\documentclass[
showpacs,preprintnumbers,amsmath,amssymb]{revtex4}
\usepackage{graphicx}% Include figure files
\begin{document}
\title{Generalized Fokker-Planck equation, Brownian motion, and ergodicity}
\author{A.V. Plyukhin}
 \affiliation{ Department of Physics and Engineering Physics,
 University of Saskatchewan, Saskatoon, SK S7N 5E2, Canada 
}

\date{\today}% It is always \today, today,
             %  but any date may be explicitly specified

\begin{abstract}
Microscopic theory of Brownian motion of a particle of mass $M$
in a bath of molecules of mass $m\ll M$ 
is considered
beyond lowest order in the mass ratio $m/M$.
The corresponding Langevin equation contains nonlinear corrections to
the dissipative force, and the generalized   
Fokker-Planck equation involves derivatives of 
order higher than two. 
These equations are derived from first principles  
with  coefficients expressed in terms of correlation functions of
microscopic force on the particle. The coefficients are  
evaluated explicitly for a generalized Rayleigh model with a finite
time of molecule-particle collisions.
In the limit of a low-density bath, we recover the results 
obtained previously  for a  model with instantaneous binary collisions. 
In general case, the equations  contain additional corrections, quadratic
in bath density, originating from a finite collision time.  
These corrections survive to order $(m/M)^2$ and  are 
found to make the  stationary distribution non-Maxwellian. 
Some relevant numerical simulations are also presented. 
\end{abstract}


\maketitle

\section{Introduction}
This paper is concerned with  the derivation and some properties of
the generalized Fokker-Planck equation (FPE) for the distribution
function $f(p,t)$ of a single stochastic variable $p$,
which  differs from the conventional FPE by 
involving $p$-derivatives  of order  higher than two,
\begin{eqnarray}
\frac{\partial f(p,t)}{\partial t}=
\sum_{n=1}^{k>2}\, \frac{\partial^n }{\partial p^n}\, c_n(p)\,f(p,t). 
\label{GFPE}
\end{eqnarray}  
Such  an equation may
appear as a result of a high-order 
truncation of  the Kramers-Moyal expansion of the Markovian master 
equation~\cite{Risken},
\begin{eqnarray}
\frac{\partial f(p,t)}{\partial t}&=&
\int dp'\Bigl\{
f(p',t) w(p'\to p)-f(p,t)w(p\to p')
\Bigl\} \label{ME}\\
&=&\sum_{n=1}^\infty\frac{1}{n!}\left(-\frac{\partial}{\partial p}\right)^n
\Bigl\{\alpha_n(p)f(p,t)\Bigr\}.\nonumber
\end{eqnarray}
In general, the naive 
truncation of this expansion is not a legitimate procedure, 
but if a problem at hand involves  a small parameter $\lambda$, 
one can approximate the expansion by finite number of terms
using an appropriate  
perturbation technique~\cite{Kampen_book}. 
In case of  Brownian motion of a heavy 
particle of mass $M$ in a thermal bath of light molecules
of mass $m$ and temperature $T$, the  appropriate small parameter is
the mass ratio $\lambda^2=m/M$. In this case, to order $\lambda^2$  
one recovers for the particle's   momentum $p$ 
the conventional second-order FPE,
\begin{eqnarray}
\frac{\partial f(p,t)}{\partial t}=
\left\{
a_1\frac{\partial }{\partial p}\, p+
a_2\frac{\partial^2 }{\partial p^2}
\right\}
\,f(p,t),
\label{FPE}
\end{eqnarray}  
 while going beyond  
order $\lambda^2$ leads to an equation
of the form (\ref{GFPE}).

In rare cases when transition rates $w$ in the master equation
(\ref{ME}) are known explicitly, 
the derivation of the generalized FPE (\ref{GFPE}) is 
fairly straightforward~\cite{Kampen_book,Kampen_paper,Alkemade,physA}, 
otherwise 
it is difficult.
A popular approach engages the  
assumption that the fluctuating force in the corresponding Langevin
equation is a Gaussian process. 
In this case  perturbation
analysis is not needed since the terms with derivatives of order higher
than two vanish identically and one arrives at the conventional
FPE (\ref{FPE}), which in  this case is exact.
However, the assumption of  
Gaussian random force, although might seem physically reasonable, 
should not be taken for granted.
% and 
%in many cases does not actually hold.
In particular, 
for a Brownian particle it is justified only to order $\lambda^2$,
while  corrections of higher orders are essentially non-Gaussian
and lead to a  FPE in the generalized form (\ref{GFPE}). 
This paper is focused on  the  generalized FPE for the   
Brownian particle's momentum $p$ to order $\lambda^4$
which 
involves $p$-derivatives up to order four.

Despite a few important 
contributions (see below), 
the theory of Brownian motion beyond the lowest approximation
did not attract much attention in the past,
perhaps because of 
the common  belief that higher order  corrections are of little importance.
However,  in recent years 
the problem has caught some new interest.
Beyond lowest order in $\lambda$ the Langevin equation 
for a Brownian particle involves nonlinear dissipative terms, and thus
corresponds to the description beyond the level of linear response
theory. One might hope that the nonlinear Langevin equation and 
the corresponding generalized FPE would
enable to capture subtle effects of the interplay of noise and 
nonlinearity, which are completely washed out when one uses the
conventional  FPE or the corresponding linear Langevin equation. 
This indeed has been demonstrated for a number of problems including  
Brownian motors~\cite{piston,Broeck1,Broeck2,Aaron} and 
barrier crossing~\cite{Kosov}.

The main difficulty related to the generalized FPE and the  corresponding
nonlinear Langevin equation  is that these equations usually  
can not be constructed on a purely phenomenological basis.
There is a substantial mathematical literature on the Langevin equation with 
nonlinear dissipation terms. However,  for processes with 
nonlinear dissipation  
thermodynamics provides no hints 
about the form of fluctuation-dissipation relations, and
little progress can be made without such relations. 
Furthermore, an attempt to go beyond 
the comfortable but artificial 
assumption of a Gaussian random
force leaves one, within a phenomenological
framework, with no clue how to
handle correlations higher than of second order.

These difficulties suggest to derive the generalized FPE
from as close to first principles as possible.
A successful example of such derivation 
is the van Kampen's method of system size 
expansion~\cite{Kampen_book} 
and in particular its application for the 
Rayleigh model of Brownian motion~\cite{Kampen_paper,Alkemade,physA}. 
In this model a Brownian particle of mass $M$
moves in one dimension
interacting with the heat bath of  
ideal gas molecules of mass $m\ll M$ and temperature $T$ through
instantaneous binary collisions. As was mentioned above,   
the relevant small parameter is
the mass ratio $\lambda^2=m/M$. 
Since the momentum of the particle $P$ is on average
$\lambda^{-1}$ times larger than that of a bath
molecule, it is convenient to work with the scaled particles momentum
$p=\lambda P$ which is of the same order as the thermal momentum of 
molecules of the bath $p_T=\sqrt{m\,k_BT}$.
The Rayleigh model is truly Markovian, and  
the  natural starting point is the master equation (\ref{ME}) for the
distribution  function $f(p,t)$, 
in which transition rates can be
readily found explicitly under the assumption of 
binary particle-molecule collisions. 
Transforming the 
Kramers-Moyal expansion  into the expansion
in powers of $\lambda$, one derives to order $\lambda^2$ 
the conventional FPE 
\begin{eqnarray}
\frac{\partial f(p,t)}{\partial t}=\lambda^2 D_2 f(p,t),
\label{FPE0}
\end{eqnarray}
where $D_2$ is a second order differential operator
\begin{eqnarray}
D_2=\gamma_0\left\{
\frac{\partial}{\partial p}\,p+
p_T^2\frac{\partial^2}{\partial p^2}
\right\},
\,\,\,\,\,\,
\gamma_0=\frac{8\nu}{\sqrt{2\pi}}\frac{p_T}{m},
\label{D2_binary}
\end{eqnarray}
$\nu$ is the number of bath molecules per unit length, 
$p_T=\sqrt{m/\beta}$ is the molecule's thermal  momentum, and
$\beta$ is the inverse temperature
$\beta=1/k_BT$.   An extension to order 
$\lambda^4$  leads to the generalized FPE in  the
form
\begin{eqnarray}
\frac{\partial f(p,t)}{\partial t}=
\Bigl\{\lambda^2 D_2+\lambda^4 D_4\Bigr\} f(p,t),
\label{VKE}
\end{eqnarray}
where the operator $D_4$ involves derivatives up to order four, 
\begin{eqnarray}
D_4=\gamma_0
\left\{
-\frac{\partial}{\partial p}\,p
+\frac{1}{6p_T^2}\,\frac{\partial}{\partial p}\,p^3
-2p_T^2\frac{\partial^2}{\partial p^2}
+\frac{3}{2}\frac{\partial^2}{\partial p^2}p^2+\frac{8p_T^2}{3}\frac{\partial^3}{\partial p^3}p
+\frac{4p_T^4}{3}\frac{\partial^4}{\partial p^4}
\right\}.
\label{D4_binary}
\end{eqnarray}
Note that terms of order $\lambda^3$ vanish due to symmetry. 
In what follows I will refer to Eq. (\ref{VKE}), 
first obtained in~\cite{Kampen_paper},
as the van Kampen equation. 

One can verify that the Maxwellian distribution
$f_M(p)=C\exp\left(-\beta\, p^2/2m\right)$  is the stationary solution
for both the  standard FPE (\ref{FPE0}) and the van Kampen equation 
(\ref{VKE}),
\begin{eqnarray}
D_2 f_M(p)=D_4 f_M(p)=0. 
\end{eqnarray}
However, in contrast to the conventional FPE (\ref{FPE0}), 
the van Kampen equation  (\ref{VKE}) does not preserve the 
positivity of the solution 
and therefore can not be an exact equation for any
stochastic process. 
Yet, as an approximation beyond lowest order
in $\lambda$, the equation (\ref{VKE}) is useful  and 
predicts a number of qualitatively new features.
For instance, while the  conventional FPE (\ref{FPE0}) gives for 
the average momentum $\langle p(t)\rangle=\int dp f(p,t) p$
the closed equation 
$\langle \dot p\rangle=-\lambda^2\gamma_0\,\langle p\rangle$, the
van Kampen equation (\ref{VKE}) predicts the coupling 
to  the third moment  $\langle p^3\rangle$:
\begin{eqnarray}
\frac{d}{dt}\langle p\rangle=-\lambda^2\gamma_0(1-\lambda^2)\,
\langle p\rangle
-\frac{1}{6}\lambda^4\gamma_0 p_T^{-2}\,\langle p^3\rangle.
\label{moment}
\end{eqnarray}
For initial
conditions $\langle p(0)\rangle=0$ and $\langle p^3(0)\rangle\ne 0$,
the FPE (\ref{FPE0}) gives $\langle p(t)\rangle=0$ for any $t>0$,
while the equation (\ref{moment}) gives nonzero average momentum 
for a time interval $t\sim \gamma_0^{-1}$. This prediction was
discussed and 
confirmed  by numerical simulation in~\cite{Aaron}.
The van Kampen equation (\ref{VKE}) has been 
recently exploited 
in the context of 
rectification of thermal fluctuations~\cite{Broeck1} and 
to study the influence of nonlinear dissipation on the Kramers 
escape rate~\cite{Kosov}.

Although proved to be useful, the van Kampen equation~(\ref{VKE})
is by no means general.
It is derived under assumptions similar to those for the Boltzmann equation,
namely that a characteristic collision time $\tau_c$ is much shorter than 
all other relevant time scales (which implies small $\lambda$), 
and that multiple collisions are negligible (small bath's density).  
It is of interest to derive a generalized  FPE from first principles 
keeping the former
assumption, but relaxing the latter.
Some aspects of this  problem were addressed already in pioneering works
on the microscopic theory of Brownian motion~\cite{pioneers} and further
developed in~\cite{Mori,Chang,Kapral}.  Perhaps the most elaborate work 
is the paper by van Kampen and Oppenheim~\cite{VKO}.
They applied the projection operator technique directly to the 
Liouville equation for the total particle-bath distribution function
and derived a generalized FPE of order $\lambda^4$.
The coefficients in the equation are expressed in terms of rather 
complicated correlation functions, and 
no attempt has been made to compare the result 
with the van Kampen equation~(\ref{VKE}). 

It was noted by Seke~\cite{Seke} that 
the projection operator method, when applied to the Liouville
equation, involves some subtlety and may be inconsistent.
The alternative way, which we shall follow in this paper, is to 
apply the projector operator technique directly to the equation of motion  
of the particle, to derive microscopically 
the Langevin equation for the particle's momentum $p$,  
and then to apply a standard 
routine~\cite{Risken}
to construct a corresponding FPE for the distribution function $f(p,t)$.

It is generally believed that the two methods of derivation, namely
``Liouville $\to$
Fokker-Planck'' and ``Equation of motion $\to$  Langevin $\to$ Fokker-Planck'',
should give the same result.
To lowest order $\lambda^2$ it is indeed the case:  
both methods lead to the standard second-order FPE (\ref{FPE}).
However, we have found recently~\cite{BP} that beyond the lowest order  
predictions of the two methods are different. 
Moreover, it was found that 
the second method 
(from Langevin to Fokker-Planck)
leads to a generalized FPE  with a rather disturbing property, namely,
its stationary solution was found non-Maxwellian. 
The intention of this paper is to follow this line in detail
to obtain a generalized FPE to order $\lambda^4$
in a complete form. 
The equation we arrive at, namely Eq.(\ref{FPE4}) in Section V,   
involves terms linear and
quadratic in the density of bath molecules $n$. For a very diluted bath
the latter can be neglected, and
the equation is reduced to the van Kampen  equation (\ref{VKE}), as expected.
The results related to the terms quadratic in $n$ are controversial.
These terms are 
absent in the van Kampen equation for a binary collisions model,  
have the structure different than
that in the van Kampen-Oppenheim approach, and  
make the stationary solution 
inconsistent with Boltzmann-Gibbs statistics.
Whether these results provide  a consistent  proof of weak 
nonergodicity of Brownian motion
or are merely 
an indication that a naive perturbation scheme does not
apply beyond the Markovian approximation     remains an open question. 
As will be discussed, a direct unambiguous verification 
of non-ergodic effects by 
numerical simulation may be a non-trivial task. 

The possibility of deviations from the Maxwellian 
distribution has been discussed in literature for a long time. It is known
that even small deviations may be important for thermally activated
processes, in particular for thermonuclear reaction 
rates in astrophysical plasma~\cite{LQ}. Several mechanisms
were proposed to justify such deviations, but in our opinion 
none of them are quite satisfactory. For instance, 
introducing a nonlinear dissipating term into the Langevin equation
and assuming that the fluctuating force is Gaussian, one generally
arrives at
a second order FPE with a non-Maxwellian stationary solution~\cite{KQ}.
However, as we already noted, nonlinear corrections to the dissipation
force for a Brownian particle is of order higher than $\lambda^2$. 
In this case
an accurate perturbation procedure leads to 
a generalized FPE of order higher than two, which  is 
a clear indication that the assumption  of Gaussian random force
is not justified beyond the lowest order.

The plan of the paper is as follows. In sections II and III, the
Mazur-Oppenheim version of the projection operator technique  
is applied to derive the nonlinear Langevin equation of order
$\lambda^4$. In this part we mostly follow the previous paper~\cite{piston1} 
making some small yet important corrections. The corresponding
generalized FPE
is constructed in Section IV and analyzed in Section V.
In Section VI we present the
results for the generalized Rayleigh model which allows analytical
evaluation of all relevant correlation functions. Also, in this
section the results of numerical simulation are discussed.
Summarizing remarks are collected in Section VII.

\section{Non-Markovian Langevin equation}
Consider a structureless Brownian particle of mass
$M$ immersed in a thermal bath comprised of  molecules of  mass $m$.
It is assumed that the mass ratio $\lambda^2=m/M$ is small and 
that the bath is initially in equilibrium at
temperature $T$. The aim of this and the next sections is to derive 
the Langevin
equation for the particle to order $\lambda^4$. As will be 
shown, such an equation involves
a nonlinear correction to the damping term, which is  
cubic in the particle's momentum.

The Hamiltonian of the system is
\begin{eqnarray}
H&=&\frac{P^2}{2M}+H_0,\\
H_0&=&\sum_i\frac{p_i^2}{2m}+U(x,X).
\end{eqnarray}
Here $x=\{x_i\}$ and $p_i$ are positions and 
momenta of bath molecules, $X$ and $P$ are those of the Brownian particle, 
$H_0$ is the Hamiltonian of the bath in the field of the Brownian
particle fixed at $X$, and the potential $U$ describes the intermolecular
interaction and the interaction  between molecules and the particle.
To simplify notations we shall consider a one-dimensional
problem. The coupling of the particle with hydrodynamic modes of 
the bath will be neglected, in which case the extension to higher
dimensions is simple.

As was already noted, it is convenient to work with 
the scaled momentum of the particle  
$p=\lambda P$,  since this quantity on
average  is expected to be of the same order as the typical momentum
of a bath molecule $p_T=\sqrt{mk_BT}=\sqrt{m/\beta}$.   
Writing the Liouville operator $\mathcal L$ in terms of $p$
has the
advantage of extracting the small parameter $\lambda$ explicitly,
\begin{eqnarray}
{\mathcal L}&=&{\mathcal L}_0+\lambda {\mathcal L}_1,\\
{\mathcal L}_0&=&\sum_i\left\{\frac{p_i}{m}\frac{\partial}{\partial x_i}+
F_i\frac{\partial}{\partial p_i}\right\},\label{L0}\\
{\mathcal L}_1&=&\frac{p}{m}\frac{\partial}{\partial X}+
F\frac{\partial}{\partial p}.\label{L1}
\end{eqnarray}
Here
$F_i=-\partial U/\partial x_i$ and $F=-\partial U/\partial X$ are 
the forces on the {\it i}-th bath molecule
and on the Brownian particle, respectively. 

The operator ${\mathcal L}_0$ corresponds to 
the Hamiltonian $H_0$ and 
governs dynamics 
of the bath in the field of the Brownian particle which is fixed at
$X$. Since $M\gg m$ one might expect that  
the force $F(t)=e^{{\mathcal L}t}F(0)$ exerted by the bath on the particle 
is close to the force on a fixed  particle (pressure), 
\begin{equation}
F(t)\approx F_0(t)\equiv e^{{\mathcal L}_0t}F(0).
\end{equation}
This intuition is implemented in
the Mazur-Oppenheim approach~\cite{MO} 
using the projection operator $\mathcal P$ 
which averages a dynamical variable $A$ over the canonical distribution
$\rho=Z^{-1}exp(-\beta H_0)$, 
for bath variables at $t=0$, 
\begin{equation}
{\mathcal P}A=\langle A\rangle \equiv\int \rho A \prod_i dx_i\, dp_i.
\end{equation}
The idea is to decompose the total force $F(t)$ on the particle 
into a zero centered fluctuating (``random'')  and 
a regular (``dissipative'') parts.
Using the operator identity
\begin{equation}
e^{{\mathcal(A+B)}t}=e^{{\mathcal A}t}+\int_0^t d\tau 
e^{{\mathcal A}(t-\tau)}{\mathcal B} e^{({\mathcal A+B})\tau},
\label{identity}
\end{equation}
with ${\mathcal A}={\mathcal L}$ and ${\mathcal B}=-{\mathcal P}{\mathcal L}$, 
one may decompose the force $F(t)=e^{\mathcal L t}F(0)$ 
as follows
\begin{eqnarray}
F(t)&=&F^\dagger(t) +\int_0^t d\tau\,\,
e^{{\mathcal L}(t-\tau)}{\mathcal P}{\mathcal L}F^\dagger(\tau),
\label{randomForce}
\end{eqnarray}
where $F^\dagger(t)=e^{{\mathcal Q}{\mathcal L}t}F$ and 
${\mathcal Q}=1-{\mathcal P}$. The factor 
${\mathcal P}{\mathcal L}F^\dagger(\tau)$ in the integral in 
Eq.~(\ref{randomForce})
can be simplified taking into account the orthogonality of ${\mathcal P}$
and ${\mathcal L}_0$ (${\mathcal P}{\mathcal L}_0=0$), and the equality 
\begin{equation}
\left\langle\frac{\partial}{\partial X} F^\dagger(t)\right\rangle=
-\beta\langle F(0)F^\dagger(t)\rangle ,
\label{gradient}
\end{equation}
which can be verified by integration by parts.
As a result, one can write 
the equation of motion for the scaled momentum 
$\dot{p}(t)=\lambda F(t)$ in the form~\cite{MO}
\begin{equation}
\frac{dp(t)}{dt}={\lambda}^2\int_0^t d\tau\,\,
e^{{\mathcal L}(t-\tau)}
\left(\frac{\partial}{\partial p}-
\frac{\beta}{m}p\right)
\langle F(0)F^\dagger(\tau)\rangle+
\lambda F^\dagger(t),
\label{EM}
\end{equation}
where $F^{\dagger}(t)$ is 
a fluctuating  force obeying
$\langle F^{\dagger}(t)\rangle=
{\mathcal P}e^{{\mathcal Q}{\mathcal L}t}F(0)=0$.

The equation (\ref{EM}) is exact but hardly instructive.
To get some progress one needs to 
expand the fluctuating 
force $F^{\dagger}(t)=e^{({\mathcal L}_0+\lambda{\mathcal
Q}{\mathcal L}_1)t}F$ in powers of $\lambda$, 
\begin{eqnarray}
F^{\dagger}(t)=F_0(t)+\lambda\,F_1(t)+\lambda^2\,F_2(t)+O(\lambda^3). 
\end{eqnarray}
Using the identity (\ref{identity}) one obtains 
\begin{eqnarray} 
F_0(t)&=&e^{\mathcal L_0t}F(0),\label{F1F2}\\
F_1(t)&=&\int_0^t \!\!dt_1\,
e^{{\mathcal L}_0(t-t_1)}{\mathcal Q}{\mathcal L}_1 F_0(t_1),\nonumber\\
F_2(t)&=&\int_0^t \!\!dt_1\, 
e^{{\mathcal L}_0(t-t_1)}{\mathcal Q}{\mathcal L}_1 
\int_0^{t_1}\!\! dt_2 \,
e^{{\mathcal L}_0(t_1-t_2)}{\mathcal Q}{\mathcal L}_1
F_0(t_2).\nonumber
\end{eqnarray}
In contrast to the pressure  $F_0(t)$, 
the higher order corrections  
$F_1(t)$ and  $F_2(t)$ depend on the particle's momentum $p$.
Seeking an equation for $p(t)$,  
one needs to extract this dependence explicitly:
\begin{eqnarray}
F_1(t)&=&\frac{p}{m}\int_0^t \!dt' 
\Bigl\{G_1(t,t')-\langle G_1(t,t')\rangle\Bigr\},\label{F12}\\
F_2(t)&=&
\left(\frac{p}{m}\right)^2\!\int_0^t\!\! dt'\!\! 
\int_0^{t'}\!\! dt'' 
\Bigl\{G_2(t,t',t'')-\langle G_2(t,t',t'')\rangle\Bigr\}\nonumber\\
&+&\frac{1}{m}\int_0^t\!\! dt'\!\! \int_0^{t'}\!\! dt'' G_0(t-t')
\Bigl\{G_1(t,t'')-\langle G_1(t,t'')\rangle\Bigr\}.\nonumber
\end{eqnarray} 
Here the functions $G_i$ are defined as follows
\begin{eqnarray}
G_0(t) &=& F_0(t),\label{Gs}\\
G_1(t,t_1) &=& \mathcal S(t-t_1)F_0(t_1),\nonumber\\
G_2(t,t_1,t_2) &=&
\mathcal S(t-t_1)\mathcal S(t_1-t_2)F_0(t_2),\nonumber
\end{eqnarray}
and  the operator $\mathcal S$ is 
\begin{equation}
\mathcal S(t_i-t_k)=e^{{\mathcal L}_0(t_i-t_k)}\frac{\partial}{\partial X}.
\end{equation}
The functions $G_i$ do not depend on the particle momentum $p$ and, 
as we shall see, all coefficients in the
$\lambda^4$-order generalized FPE can be expressed in terms of correlation
functions of $G_0$, $G_1$, and $G_2$.

Let us now return to the exact equation of motion Eq.(\ref{EM}) and 
expand the correlation $\langle F(0)F^\dagger (t) \rangle$ 
to order $\lambda^2$, 
\begin{eqnarray}
\langle F(0)F^\dagger (t)\rangle=
\langle F(0)F_0(t)\rangle+
\lambda^2\left\langle F(0) F_2(t)\right\rangle.
\end{eqnarray}
Note that $\langle F(0)F_1(t)\rangle=0$ due to symmetry.
Using (\ref{F12}), one can write the above equation in the following form
\begin{eqnarray}
\langle F(0)F^{\dagger}(t)\rangle=C_0(t)+
\lambda^2\left\{
\left(\frac{p}{m}\right)^2C_1(t)+\frac{1}{m}\,C_2(t)\right\},
\label{kernel_exp}
\end{eqnarray}
where
\begin{eqnarray}
&&C_0(t)=\langle\!\langle G_0(0)G_0(t)\rangle\!\rangle,\label{Cs}\\
&&C_1(t)=\!\int_0^t\!\! dt'\!\!\int_0^{t'}\!\! dt''
\langle\!\langle G_0(0)G_2(t,t',t'')\rangle\!\rangle,\nonumber\\
&&C_2(t)=\!\int_0^t\!\! dt'\!\!\int_0^{t'}\!\! dt''
\langle\!\langle G_0(0)G_0(t-t')G_1(t,t'')\rangle\!\rangle\nonumber.
\end{eqnarray}

In the above equations, cumulants   
$\langle\!\langle A_1A_2\cdots A_k\rangle\!\rangle$ are defined 
in a usual way, i.e. as
a part of the correlation $\langle A_1A_2\cdots A_k\rangle$
which can not be reduced to the product of correlations of lower
order. We shall need correlations and cumulants up to order four,
which are related to each other as follows:   
\begin{eqnarray} 
\langle A\rangle&=&\langle\!\langle A\rangle\!\rangle,\\
\langle A_1A_2\rangle&=& \langle A_1\rangle\langle A_2\rangle+
\langle\!\langle A_1A_2\rangle\!\rangle,\nonumber\\
\langle A_1A_2A_3\rangle& =& \langle A_1\rangle\langle A_2\rangle
\langle A_3\rangle+
\langle A_1\rangle \langle\!\langle A_2A_3\rangle\!\rangle
+\langle A_2\rangle \langle\!\langle A_3A_1\rangle\!\rangle+
\langle A_3\rangle \langle\!\langle A_2A_1\rangle\!\rangle+
\langle\!\langle A_1A_2A_3\rangle\!\rangle,\nonumber\\ 
\langle A_1A_2A_3A_4\rangle& =& \langle A_1\rangle\langle A_2\rangle
\langle A_3\rangle\langle A_4\rangle\nonumber\\
&&+\langle A_1\rangle \langle\!\langle A_2A_3A_4\rangle\!\rangle
+\langle A_2\rangle \langle\!\langle A_1A_3A_4\rangle\!\rangle+
\langle A_3\rangle \langle\!\langle A_1A_2A_4\rangle\!\rangle+
\langle A_4\rangle \langle\!\langle
A_1A_2A_3\rangle\!\rangle\nonumber \\
&&+\langle\!\langle A_1A_2\rangle\!\rangle
\langle\!\langle A_3A_4\rangle\!\rangle+
\langle\!\langle A_1A_3\rangle\!\rangle 
\langle\!\langle A_2A_4\rangle\!\rangle+
\langle\!\langle A_1A_4\rangle\!\rangle
\langle\!\langle A_2A_3\rangle\!\rangle+
\langle\!\langle A_1A_2A_3A_4\rangle\!\rangle.
\nonumber 
\label{cumexp}
\end{eqnarray}
The important property of cumulants of functions $G_i$ 
is that 
they are linear in the concentration of bath molecules $n$. 
\begin{eqnarray}
\langle\!\langle G_iG_j\cdots G_k\rangle\!\rangle\sim n.
\end{eqnarray}
This may be demonstrated noticing that
$G_i$ is a linear functional of cumulants for
the density of bath particles $N(z,t)=\sum_i\delta(z-z_i(t))$,
where $z_i$ denotes
the coordinate-momentum pair $(x_i,p_i)$ of a bath particle.
In turn, one can observe that cumulants 
$\langle\!\langle N(z_1,t_1)N(z_2,t_2)...N(z_k,t_k)\rangle\!\rangle$
of any order $k$
depend  linearly on the concentration of bath molecules $n$.  
For instance,
in the expression for the product
$N(z_1,t_1)N(z_2,t_2)$ one can write the double sum as 
$\sum_{i,j}=\sum_{i\ne j}+\sum_{i=j}$ which gives
\begin{eqnarray}
\langle N(z_1,t_1)N(z_2,t_2)\rangle=
\langle N(z_1,t_1)\rangle\langle N(z_2,t_2)\rangle+
\sum_{i}\langle\delta(z_1-z_i(t_1))\delta(z_2-z_i(t_2))\rangle.
\end{eqnarray} 
Here the second term on the right side 
is by definition the cumulant
$\langle\!\langle N(z_1,t_1)N(z_2,t_2)\rangle\!\rangle$
and is  obviously linear in $n$.

Substituting the expansion (\ref{kernel_exp}) into the equation of
motion (\ref{EM}) one arrives at
a nonlinear and non-Markovian Langevin equation of order 
$\lambda^4$,
\begin{eqnarray}
\frac{dp(t)}{dt}= -
{\lambda}^2\int_0^t \!d\tau\,
M_1(\tau) p(t-\tau)-{\lambda}^4\int_0^t \!d\tau\,
M_2(\tau) p^3(t-\tau)+\lambda F(t).
\label{NMLE}
\end{eqnarray}  
Here the memory kernels are
\begin{eqnarray}
M_1(t)&=&\frac{\beta}{m}C_0(t)-\lambda^2\frac{2}{m^2} C_1(t)+
\lambda^2\frac{\beta}{m^2}C_2(t),\label{M1}\\
M_2(t)&=&\frac{\beta}{m^3}C_1(t),\nonumber
% \label{M2}
\end{eqnarray}
and the fluctuating force $\lambda F(t)$ involves three components,
\begin{eqnarray}
F(t)=F_0(t)+\lambda F_1(t)+\lambda^2 F_2(t),
\end{eqnarray}
where $F_0(t)$ is the fluctuating pressure, and $F_1(t)$ and $F_2(t)$
are defined by Eqs.(\ref{F12}). The zero-mean term 
$\lambda^3 F_3(t)$ is discarded  
because  
to order $\lambda^4$ it does not 
contribute to correlations of the  fluctuating
force $\lambda F(t)$.

\section{Markovian Langevin equation}
The relative importance of memory effects described by 
the non-Markovian Langevin equation (\ref{NMLE}) depends on how fast
the particle's momentum  $p(t)$ evolves on the 
time scale $\tau_c$
for the decay of the memory kernels $M_1(t)$ and $M_2(t)$. 
In what follows we shall assume that the characteristic time 
$\tau_c$
does exist. This assumption is not satisfied, 
for example, for the Rubin's model 
(a heavy impurity embedded in the harmonic chain)
where $M_2(t)$ is identically zero, and $M_1(t)$ 
decays with time as a power law. It is known, however, that
in many cases long-tail effects are indeed negligible for sufficiently
small $\lambda$, although the justification may require rather subtle  
argument~\cite{Kapral,Mazo}

The memory kernels $M_i(t)$ are expressed in terms of correlation functions
$C_i(t)$ which do not depend on $\lambda$. Then 
the characteristic decay time of the kernels does not depend  on $\lambda$
either, $\tau_c\sim\lambda^0$.  On the other hand,
as follows from Eq.(\ref{NMLE}), the  characteristic 
time for relaxation of the particle's momentum $\tau_p$ is of
order $\lambda^{-2}$ and thus expected to be much longer than 
$\tau_c$. 
This suggests that the non-Markovian equation
(\ref{NMLE}) can be expanded in powers of $\lambda$ 
about its Markovian limit. 

One way to make such an expansion 
is  to use in Eq. (\ref{NMLE}) the following substitution~\cite{MO}
\begin{eqnarray}
p^n(t-\tau)=p^n(t)-\int_{t-\tau}^t dt'\frac{d}{dt'}{p^n}(t').
\label{trick}
\end{eqnarray}
The main contributions to the integrals in Eq.(\ref{NMLE})
come from the region $\tau\le\tau_c\sim\lambda^0$. For such $\tau$ 
the integral term in the right-hand side
of Eq.(\ref{trick})  is of order $\dot p\sim\lambda$. Then  the  
nonlinear dissipative term in the Langevin equation (\ref{NMLE}) can 
be written in the
local form 
\begin{eqnarray}
-{\lambda}^4\int_0^t d\tau\,\,
M_2(\tau)\, p^3(t-\tau)=
-{\lambda}^4\,p^3(t)\int_0^t d\tau\,\,
M_2(\tau)+O(\lambda^5).
\end{eqnarray}
To order $\lambda^4$ this justifies the Markovian 
ansatz for the nonlinear dissipative term
\begin{eqnarray}
-{\lambda}^4\int_0^t d\tau\,\,
M_2(\tau)\, p^3(t-\tau)\to
-{\lambda}^4\,p^3(t)\int_0^\infty d\tau\,\,
M_2(\tau).
\label{Mapprox}
\end{eqnarray}
Here the upper integration limit in the right-hand side is taken to infinity
since we restrict ourselves to 
the coarse-grain description on the time scale 
much longer than the characteristic  time for bath fluctuations,
$t\gg\tau_c$.

The same argument for the linear dissipative term gives 
\begin{eqnarray}
-{\lambda}^2\int_0^t d\tau\,\,
M_1(\tau)\, p(t-\tau)=
-{\lambda}^2\,p(t)\int_0^t d\tau\,\,
M_0(\tau)+O(\lambda^3).
\end{eqnarray}
In contrast to the nonlinear dissipation term, 
the Markovian approximation 
\begin{eqnarray}
-{\lambda}^2\int_0^t d\tau\,\,
M_1(\tau)\, p(t-\tau)\to
-{\lambda}^2\,p(t)\int_0^\infty d\tau\,\,
M_0(\tau)
\end{eqnarray}
can be applied for the linear dissipative force only in lowest order
$\lambda^2$, in which case one recovers
the conventional linear Langevin equation 
\begin{eqnarray}
\frac{dp(t)}{dt}= -
{\lambda}^2\,\gamma_0 \, p(t)+\lambda F_0(t),
\label{LE0}
\end{eqnarray}  
with the pressure $F_0(t)$ as a fluctuating force and 
the dissipation constant 
\begin{eqnarray}
\gamma_0=\frac{\beta}{m}\int_0^\infty dt\, C_0(t).
\end{eqnarray}

Let us now derive a local form for  the linear dissipative term to order
$\lambda^4$. Using again 
the substitution (\ref{trick}),
one can write  the linear term as a  
local expression plus a correction term 
$\Delta(t)$
\begin{eqnarray}
-{\lambda}^2\!\!\int_0^t\! d\tau\,
M_1(\tau)\, p(t-\tau)=
-{\lambda}^2 p(t)\!\int_0^t\! d\tau\,\,
M_1(\tau)+\Delta (t).
\label{ansatz}
\end{eqnarray}
The correction has a form 
\begin{eqnarray}
\Delta(t)={\lambda}^2\,\int_0^t \!d\tau\,
M_1(\tau)\int_{t-\tau}^t\!d\tau'\,\dot{p}(\tau').
\end{eqnarray}  
Here the essential integration range is of order 
$\tau_c\sim\lambda^0$, so that the correction term
$\Delta\sim \lambda^2\dot p\sim\lambda^3$.
Using the linear Langevin equation (\ref{LE0}),  one can write $\Delta$
as follows
\begin{eqnarray}
\Delta(t)=-\lambda^4\gamma^\star p(t)+\lambda^3F^\star(t).
\label{delta}
\end{eqnarray} 
Here the first term $-\lambda^4\gamma^\star p$ with 
\begin{eqnarray}
\gamma^\star=\left(\frac{\beta}{m}\right)^2\int_0^\infty \!\!dt\,C_0(t)\,
\int_0^\infty \!\!dt\,C_0(t)\,t
\label{gamma_star}
\end{eqnarray}
is a correction to the local 
linear dissipative force, while 
the second term
\begin{eqnarray}
F^\star(t)=\frac{\beta}{m}\,\int_0^\infty \!\!d\tau\,
C_0(\tau)\int_{t-\tau}^t\!\!dt'
 F_0(t')
\label{Fstar}
\end{eqnarray}
can be considered as an additional contribution to the fluctuating  force.
The above expressions for $\gamma^\star$ and 
$F^\star(t)$ are valid for time scale $t\gg \tau_c$. One can show that
in this case
$F^\star(t)$ is a stationary process.  Note that the contribution
$F^\star(t)$ was overlooked in~\cite{piston1}.

The result (\ref{delta}) may be  also obtained in a more direct  
way expanding $p(t-\tau)$ about $t$:
\begin{eqnarray}
\lambda^2\int_0^t\! d\tau\,
M_1(\tau)\, p(t-\tau)=&&\lambda^2p(t)\!\int_0^t\! d\tau\,\,M_1(\tau)\\
&&-\lambda^2\dot{p}(t)\!\int_0^t\! d\tau\,\,M_1(\tau)\tau
+\frac{1}{2!}\lambda^2\ddot{p}(t)\!\int_0^t\!
d\tau\,\,M_1(\tau)\tau^2
+\cdots
\nonumber
\end{eqnarray}
To evaluate this expression to order $\lambda^4$ one needs derivatives
$p^{(n)}(t)$ to order $\lambda^2$. 
The first derivative is given by
the linear Langevin equation (\ref{LE0}), while for
derivatives of higher order Eq.(\ref{LE0}) gives:
$p^{(n)}(t)=\lambda F_0^{(n-1)}(t)+O(\lambda^3)$.
Then the above expansion can be  transformed into the form
(\ref{ansatz}) with the fluctuating term
\begin{eqnarray}
\lambda^3F^\star(t)=\lambda^3\int_0^t d\tau M_1(\tau) 
\left\{F_0(t)\tau-\frac{1}{2!}\dot{F}_0(t)\tau^2+\cdots\right\}.
\end{eqnarray}
Recalling that $M_1(t)=\frac{\beta}{m}C_0(t)+O(\lambda^2)$ and noticing that
\begin{eqnarray}
F_0(t)\tau-\frac{1}{2!}\dot{F}_0(t)\tau^2+\cdots=
\int_0^\tau dt'F_0(t-t')=\int_{t-\tau}^t dt'F_0(t')
\end{eqnarray}
one recovers $F^\star(t)$ in the form (\ref{Fstar}).

The above results allow us to write the Langevin equation  to order 
$\lambda^4$
in a local form as follows
\begin{eqnarray}
\frac{dp(t)}{dt}=
-\lambda^2\gamma_1\,p(t)-{\lambda}^4\gamma_2\,p^3(t)-\lambda\, \xi(t).
\label{MLE}
\end{eqnarray}  
Here the fluctuating force is 
\begin{eqnarray}
\xi(t)=
F_0(t)+\lambda F_1(t)+\lambda^2 F_2(t)+\lambda^2 F^\star(t),
\end{eqnarray}
and the dissipation coefficients are 
\begin{eqnarray}
\gamma_1&=&\gamma_{0}+\lambda^2\delta\gamma+\lambda^2\gamma^\star,
\label{gammas1}\\
\gamma_{0}&=&\frac{\beta}{m}\int_0^\infty dt \,C_0(t),\nonumber\\
\delta\gamma&=&-\frac{2}{m^2}\int_0^\infty dt\,C_1(t)+
\frac{\beta}{m^2}\int_0^\infty dt\,C_2(t),\nonumber\\
\gamma^\star&=&\left(\frac{\beta}{m}\right)^2\int_0^\infty \!\!dt\,C_0(t)\,
\int_0^\infty \!\!dt\,C_0(t)\,t,\nonumber\\
\gamma_2&=&\frac{\beta}{m^3}\int_0^\infty dt \,C_1(t).\nonumber
\end{eqnarray}  
Recall that the correlation functions $C_i(t)$ are defined by Eqs.(\ref{Cs}).

\section{From Langevin  to Fokker-Planck }
The aim of this section is to construct 
a Fokker-Planck equation for the distribution 
function $f(p,t)$ corresponding to the Langevin equation (\ref{MLE}).
We shall mostly follow the standard procedure~\cite{Risken}, but without 
conventional assumption that the fluctuating force is delta-correlated.
%This requires some care, so we shall discussed 
%the procedure in some detail. 

The first step is to 
assume that $f(p,t)$ obeys a Markovian master equation
\begin{eqnarray}
\frac{\partial f(p,t)}{\partial t}=
\int dp'\Bigl\{
f(p',t) w(p'\to p)-f(p,t)w(p\to p')
\Bigl\}.
\label{master}
\end{eqnarray}
Expressing the transition rates 
$w(p_1\to p_2)$ as
a function of the initial state $p_1$ and the transition length
$\Delta p=p_2-p_1$, $w(p_1\to p_2)=w(p_1|\Delta p)$, 
the master equation can be written in the form
\begin{eqnarray}
\frac{\partial f(p,t)}{\partial t}
&=&\int d(\Delta p)\Bigl\{
f(p-\Delta p,t)w(p-\Delta p|\Delta p)-f(p,t)w(p|\Delta p)\Bigr\}\nonumber\\
&=&\int d(\Delta p)\Bigl\{\Psi(p-\Delta p,\Delta p)-
\Psi(p,\Delta p)\Bigr\},
\label{aux001}
\end{eqnarray}
where $\Psi(p,\Delta p)=f(p,t)w(p|\Delta p)$.
Next, the expansion 
\begin{eqnarray}
\Psi(p-\Delta p,\Delta p)=\Psi(p, \Delta p)+
\sum_{n=1}^{\infty}
\frac{1}{n!}\left(-\Delta p\frac{\partial}{\partial p}\right)^n
\Psi(p,\Delta p)
\end{eqnarray}
transforms the master equation into the Kramers-Moyal form
\begin{eqnarray}
\frac{\partial f(p,t)}{\partial t}=
\sum_{n=1}^{\infty}\frac{1}{n!}\left(-\frac{\partial}{\partial p}\right)^n
\Bigl\{\alpha_n(p)f(p,t)\Bigr\}
\label{KME}
\end{eqnarray}
with coefficients $\alpha_n$ given by
\begin{eqnarray}
\alpha_n(p)=\int d(\Delta p)(\Delta p)^n w(p|\Delta p).
\label{alpha1}
\end{eqnarray}

Since transition rates $w$ are usually unknown,  
it is more convenient to work with another
representation for $\alpha_n$. Namely,
expressing $w$
in terms of the transition probability $T(p,t\,|\,p',t+\tau)$
\begin{eqnarray}
w(p\to p')=\lim_{\tau\to 0}\,\frac{1}{\tau}\,T(p,t\,|\,p',t+\tau)
\end{eqnarray}
one can write the expression
(\ref{alpha1}) for $\alpha_n$ in the form 
\begin{eqnarray}
\alpha_n(V)=\lim_{\tau\to 0}\,\frac{1}{\tau}\,
\int dp'(p'-p)^nT(p,t\,|\,p',t+\tau),\nonumber
\end{eqnarray}
or
\begin{eqnarray}
\alpha_n(p)=\lim_{\tau\to 0}\,\frac{1}{\tau}\,
\Bigl\langle[p(t+\tau)-p(t)]^n\Bigr\rangle.
\label{alphas0}
\end{eqnarray} 
This expression may be  evaluated integrating the Langevin
equation for $p(t)$, but this step  needs care.
Recall that the Langevin equation (\ref{MLE}) corresponds to 
a coarse-grained description with a time resolution
much longer than correlation time of the random force $\tau_c$ 
but much shorter than
the characteristic time for the relaxation of the particle's 
momentum $\tau_p$.  Then in the above expressions 
the limit $\tau\to 0$ should be understood as 
$\tau_c\ll\tau\ll \tau_p$. 
The moments 
$\Bigl\langle [p(t+\tau)-p(\tau)]^n\Bigr\rangle$ in Eq. (\ref{alphas0})
must be first evaluated
in the limit $\tau\gg\tau_c$, and only after that the formal operation
$\lim_{\tau\to 0}\frac{1}{\tau}(\cdots)$ must be applied:
\begin{eqnarray}
\alpha_n(p)=\lim_{\tau\to 0}\frac{1}{\tau}
\left\{
\lim_{\tau\gg\tau_c}
\Bigl\langle [p(t+\tau)-p(t)]^n\Bigr\rangle\right\}.
\label{alphas}
\end{eqnarray} 
In what follows, the coarse-grained limit $\tau\gg \tau_c$
for brevity will not be indicated.
%\begin{eqnarray}
%\lim_{\tau\to 0}\frac{1}{\tau} A(\tau)\equiv
%\lim_{\tau\to 0}\frac{1}{\tau}\Bigl\{\lim_{\tau\gg\tau_c} A(\tau)\Bigr\}
%\end{eqnarray} 
%In other words, it is assumed that for $\tau_c\ll \tau \ll\tau_p$
%the expression 
%\begin{eqnarray}
%\Bigl\langle(p(t+\tau)-p(t))^n\Bigr\rangle=a+b\tau+c\tau^2+\cdots
%\end{eqnarray}  
%behave asymptotically as $b\tau$, and $\alpha_n$ is identified  with $b$.

Integrating the  Langevin equation (\ref{MLE})
for $\tau_c\ll \tau \ll\tau_p$
\begin{eqnarray}
p(t+\tau)-p(t)\approx -[\lambda^2\gamma_1p(t)+\lambda^4\gamma_2p^3(t)]\tau
+\lambda\int_t^{t+\tau}dt'\xi(t')
\end{eqnarray} 
and recalling that $\xi(t)$ is a stationary process for $t\gg\tau_c$, 
one obtains from (\ref{alphas}):
\begin{eqnarray}
\alpha_1&=&-\lambda^2\,\gamma_1\,p-\lambda^4\,\gamma_2\,p^3,
\label{alphas_3}\\
\alpha_2&=&\lambda^2\lim_{\tau\to 0}
\frac{1}{\tau}
\int_0^\tau\!\! dt_1
\int_0^\tau\!\! dt_2\,\langle \xi(t_1)\xi(t_2)\rangle
,\nonumber\\
\alpha_3&=&\lambda^3\lim_{\tau\to 0}
\frac{1}{\tau}
\int_0^\tau\!\! dt_1\!
\int_0^\tau\!\!dt_2\!\int_0^\tau\! dt_3\,
\langle \xi(t_1)\xi(t_2)\xi(t_3)\rangle,\nonumber\\
\alpha_4&=&\lambda^4\lim_{\tau\to 0}
\frac{1}{\tau}
\int_0^\tau\!\! dt_1\!
\int_0^\tau\!\! dt_2\!\int_0^\tau\!\! dt_3\!\int_0^\tau \!\!dt_4
\langle \xi(t_1)\xi(t_2)\xi(t_3)\xi(t_4)\rangle.
\nonumber
\end{eqnarray}
In these expressions the integrals must be taken in the 
coarse-grained limit $\tau\gg\tau_c$.

The next step is to substitute in these expressions the fluctuating  force
$\xi=F_0+\lambda F_1+\lambda^2 F_2+\lambda^2\,F^\star$
retaining enough terms  to get $\alpha_n$ to order $\lambda^4$.
According to (\ref{alphas_3}), 
the expression for $\alpha_2$ requires 
the correlation $\langle \xi\xi\rangle$ to order
$\lambda^2$,
\begin{eqnarray}
\langle \xi(t_1)\xi(t_2)\rangle&=&
\langle F_0(t_1)F_0(t_2)\rangle+\lambda^2\langle
F_1(t_1)F_1(t_2)\rangle
+\lambda^2\langle F_0(t_1)F_0^\star(t_2)\rangle+
\lambda^2\langle F_0(t_2)F_0^\star(t_1)\rangle\nonumber\\
&&+\lambda^2\langle F_0(t_1)F_2(t_2)\rangle+
\lambda^2\langle F_0(t_2)F_2(t_1)\rangle,\label{correlations}
\end{eqnarray}
$\alpha_3$ requires 
the correlation $\langle \xi\xi\xi\rangle$ to order
$\lambda$,
\begin{eqnarray}
\!\!\!\!\!\!\!\!\!\!\!\!\!
\langle \xi(t_1)\xi(t_2)\xi(t_3)\rangle&\!\!=\!\!&
\lambda \langle F_0(t_1)F_0(t_2)F_1(t_3)\rangle
\!+\!
\lambda \langle F_0(t_1)F_1(t_2)F_0(t_3)\rangle
\!+\!
\lambda \langle F_1(t_1)F_0(t_2)F_0(t_3)\rangle,
\end{eqnarray}
and $\alpha_4$ requires 
the correlation $\langle \xi\xi\xi\rangle$ to order
$\lambda^0$,
\begin{eqnarray}
\langle \xi(t_1)\xi(t_2)\xi(t_3)\xi(t_4)\rangle&=&
\langle F_0(t_1)F_0(t_2)F_0(t_3)F_0(t_4)\rangle.
\end{eqnarray}
To extract the dependence on $p$ one has to 
express $F_1$  and $F_2$ in terms
of $p$-independent functions $G_i(t)$, see Eq. (\ref{F12}).
Then the correlation functions take the forms:
\begin{eqnarray}
&&\!\!\!\!\!\!\!\langle \xi(t_1)\xi(t_2)\rangle=
C_0(t_1,t_2)
+\lambda^2\,m^{-1}
\Bigl\{C_2(t_1,t_2)+C_2(t_2,t_1)\Bigr\}
\label{xi2}\\
&&+\lambda^2\,\left(\frac{p}{m}\right)^2\,
\Bigl\{
C_1(t_1,t_2)+C_1(t_2,t_1)+C_3(t_2,t_1)\Bigr\}
+\lambda^2\,\frac{\beta}{m}\, \Bigl\{C_6(t_1,t_2)+C_6(t_2,t_1)\Bigr\},
\nonumber\\
&&\!\!\!\!\!\!\!\langle \xi(t_1)\xi(t_2)\xi(t_3)\rangle=
\lambda\,\frac{p}{m}
\Big\{
C_4(t_1,t_2,t_3)+C_4(t_1,t_3,t_2)+C_4(t_3,t_2,t_1)
\Bigr\},\nonumber\\
&&\!\!\!\!\!\!\!\langle \xi(t_1)\xi(t_2)\xi(t_3)\xi(t_4)\rangle=
C_0(t_1,t_2)C_0(t_3,t_4)
+C_0(t_1,t_3)C_0(t_2, t_4)+C_0(t_1,t_4)C_0(t_2,t_3)
+C_5(t_1,t_2,t_3,t_4).
\nonumber
\end{eqnarray}
Here the functions $C_i$ are
defined  as follows
\begin{eqnarray}
&&C_0(t_1,t_2)=
\langle\!\langle G_0(t_1)G_0(t_2)\rangle\!\rangle,
\label{Cs2}\\
&&C_1(t_1,t_2)=\int_0^{t_2}\!\!dt'\!\int_0^{t'}\!\! dt''
\langle\!\langle G_0(t_1)G_2(t_2,t',t'')\rangle\!\rangle,\nonumber\\
&&C_2(t_1,t_2)=
\int_0^{t_2}\!\! dt'\!\int_0^{t'}\!\! dt''
\langle\!\langle
G_0(t_1)G_0(t_2-t')G_1(t_2,t'')\rangle\!\rangle,\nonumber\\
&&C_3(t_1,t_2)=\int_0^{t_1}\!\! dt'\!\int_0^{t_2}\!\! dt''
\langle\!\langle G_1(t_1,t')G_1(t_2,t'')\rangle\!\rangle,\nonumber\\
&&C_4(t_1,t_2,t_3)=\int_0^{t_3}\!dt'
\langle\!\langle G_0(t_1)G_0(t_2)G_1(t_3,t')\rangle\!\rangle,\nonumber\\
&&C_5(t_1,t_2,t_3,t_4)=
\langle\!\langle
G_0(t_1)G_0(t_2)G_0(t_3)G_0(t_4)\rangle\!\rangle,\nonumber\\
&&C_6(t_1,t_2)=\int_0^{t_2}\!\!dt'\,C_0(t')\int_{t_2-t'}^{t_2}\!\! dt''\,
C_0(t_1, t'').\nonumber
\end{eqnarray}
%and
%\begin{eqnarray}
%C_6(t_1,t_2)&=&\int_0^{t_2}dt'\,C_0(t')\int_{t_2-t'}^{t_2} dt''\,
%C_0(t_1,t'')\label{C6}.
%\end{eqnarray}
Note that 
the two variable function
$C_0(t_1,t_2)=\langle\!\langle G_0(t_1)G_0(t_2)\rangle\!\rangle$ defined
above, and 
the single variable function  
$C_0(t)=\langle\!\langle G_0(0)G_0(t)\rangle\!\rangle$
which we used in the previous sections, are related as
$C_0(t_1,t_2)=C_0(|t_1-t_2|)$.

Substitution of correlations (\ref{xi2}) into (\ref{alphas_3})
leads to the following results
\begin{eqnarray}
\alpha_1&=&-\lambda^2\gamma_1\,p-\lambda^4\gamma_2\,p^3,
%-\lambda^2(\gamma_{0}+\lambda^2\delta\gamma+\lambda^2\gamma^\star)
%\,p-\lambda^4\gamma_2\,p^3,
\label{alphas_5}\\
\alpha_2&=&
\lambda^2 \gamma_{3}+\lambda^4 \left(\frac{p}{m}\right)^2 \gamma_{4}
+\lambda^4 \frac{1}{m} \gamma_{5}+
\lambda^4 \frac{\beta}{m} \gamma_{6},\nonumber\\
\alpha_3&=&\lambda^4\,\frac{p}{m} \gamma_{7},\nonumber\\
\alpha_4&=&\lambda^4\,\gamma_8.\nonumber
\end{eqnarray}
Here the dissipation coefficients $\gamma_1$ and $\gamma_2$, 
are defined above by equations (\ref{gammas1}),
\begin{eqnarray}
&&\gamma_{3}=\lim_{\tau\to 0}
\frac{1}{\tau}\,\int_0^\tau \!\! dt_1\!\! \int_0^\tau \!\!dt_2\, 
C_0(t_1, t_2),
\label{gammas2}\\
&&\gamma_{4}=\lim_{\tau\to 0}
\frac{1}{\tau}\,\int_0^\tau \!\! dt_1 \!\!\int_0^\tau \!\!dt_2\, 
\Bigl\{
2C_1(t_1,t_2)+C_3(t_1,t_2)
\Bigr\},\nonumber\\
&&\gamma_{5}=\lim_{\tau\to 0}
\frac{1}{\tau}\,\int_0^\tau  \!\!dt_1\!\! \int_0^\tau \!\!dt_2\, 
2C_2(t_1,t_2),\nonumber\\
&&\gamma_{6}=\lim_{\tau\to 0}
\frac{1}{\tau}\,\int_0^\tau \!\!dt_1 \!\!\int_0^\tau \!\!dt_2\, 
2C_6(t_1,t_2),\nonumber\\
&&\gamma_{7}=\lim_{\tau\to 0}
\frac{1}{\tau}\,\int_0^\tau \!\!dt_1 \!\!\int_0^\tau \!\!dt_2\!\!
\int_0^\tau \!\!dt_3\, 
3C_4(t_1,t_2,t_3),\nonumber\\
&&\gamma_{8}=\lim_{\tau\to 0}
\frac{1}{\tau}\,\int_0^\tau \!\!dt_1\!\! \int_0^\tau \!\!dt_2\!\!
\int_0^\tau \!\!dt_3\!\!\int_0^\tau \!\!dt_4\, C_5(t_1,t_2,t_3,t_4).\nonumber
\end{eqnarray}
Recall again that in these formulas
the integrals must be evaluated in 
the coarse-grained limit $\tau\gg \tau_c$. 

Note that
in the formula for $\gamma_8$ we have discarded the terms involving  
the products of cumulants,  like 
$\langle\!\langle  F_0(t_1)F_0(t_2)\rangle\!\rangle 
\langle\!\langle F_0(t_3)F_0(t_4)\rangle\!\rangle $.
Such products  
depend on two time differences, and  
the corresponding contributions to 
the four-dimensional time integral in the expression for 
$\alpha_4$ are quadratic in $\tau$. They therefore
vanish when the operation
$\lim_{\tau\to 0}\frac{1}{\tau}(\cdots)$ is applied.

\section{Generalized Fokker-Planck equation}
Substituting the results (\ref{alphas_5}) for $\alpha_n$ 
into the Kramers-Moyal expansion (\ref{KME}) one  arrives at
the Fokker-Planck equation of order $\lambda^4$ in the following 
form 
\begin{eqnarray}
\frac{\partial f(p,t)}{\partial t}=
\Bigl\{\lambda^2 D_2+\lambda^4 D_4+\lambda^4 D^\star_2
\Bigr\} f(p,t).
\label{FPE4}
\end{eqnarray}
Here the differential operators $D_2$ and  $D_4$ have the same structure 
as for  the van Kampen equation (\ref{VKE}),
\begin{eqnarray}
D_2&=&a_1\,\frac{\partial}{\partial p}\,p+
a_2\,\frac{\partial^2}{\partial p^2},\\
D_4&=&b_1\frac{\partial}{\partial p}\,p+
b_2\frac{\partial}{\partial p}\,p^3
+b_3\frac{\partial^2}{\partial p^2}+
b_4\frac{\partial^2}{\partial p^2}p^2
+b_5\frac{\partial^3}{\partial p^3}p+
b_6\frac{\partial^4}{\partial p^4}.
\end{eqnarray}
The difference with 
the van Kampen equation  is the presence of 
the operator $D_2^\star$, originating from non-Markovian corrections
$\gamma^\star$ and $F^\star$. It has the same structure as $D_2$
\begin{eqnarray}
D_2^\star=c_1\,\frac{\partial}{\partial p}\,p+
c_2\,\frac{\partial^2}{\partial p^2},
\end{eqnarray}
but as we shall see, scales differently with the bath density $n$.

For the operator $D_2$ the coefficients are 
$a_1=\gamma_{0}$ and $a_2=\frac{1}{2}\,\gamma_{3}$,
\begin{eqnarray}
&&a_1=\frac{\beta}{m}\int_0^\infty \!\!dt\, C_0(t),\label{a1}\\
&&a_2=\frac{1}{2\tau}\,\int_0^\tau \!\!dt_1\!\! \int_0^\tau\!\! dt_2\, 
C_0(t_1, t_2).\nonumber
\end{eqnarray}
The expression for $a_2$ may be simplified
recalling that $C_0(t_1,t_2)=C_0(|t_1-t_2|)$ and using the
coarse-grained limit $\tau\gg\tau_c$:
\begin{eqnarray}
a_2=
\frac{1}{\tau}\int_0^\tau \!dt_2\!\int_0^{t_2} \!dt_1\,C_0(t_1,t_2)=
\frac{1}{\tau}\int_0^\tau \!dt_2 (\tau-t_2)\,C_0(t_2)\to
\int_0^\infty \!dt \,C_0(t).\label{a2}
\end{eqnarray} 
This gives the conventional relation 
\begin{eqnarray}
\frac{a_1}{a_2}=\frac{\beta}{m},
\end{eqnarray}
which guarantees that the Maxwellian distribution
$f_M(p)=C\exp(-\beta p^2/2m)$ is stationary for the operator $D_2$:
$D_2\,f_M(p)=0$.

For the operator $D_4$ the coefficients are
$b_1=\delta\gamma$, $b_2=\gamma_2$,
$b_3=\gamma_{4}/2m^2$,
$b_4=\gamma_{5}/2m$, 
$b_5=\gamma_7/3!m$, and 
$b_6=\gamma_8/4!$ 
Using the results (\ref{gammas1}) and  (\ref{gammas2}) for $\gamma_i$, 
one can write
the coefficients $b_i$ in terms of correlation functions 
as follows:
\begin{eqnarray}
&&b_1=
\frac{2}{m^2}\int_0^\infty \!\!dt\,C_1(t)+
\frac{\beta}{m^2}\int_0^\infty \!\!dt\,C_2(t),
\label{cc}\\
&&b_2=
\frac{\beta}{m^3}\int_0^\infty \!\!dt \,C_1(t),\nonumber\\
&&b_3=
\frac{1}{2m^2}\lim_{\tau\to 0}\frac{1}{\tau}\,
\int_0^\tau \!\!dt_1 \!\!\int_0^\tau \!\!dt_2\, 
\Bigl\{
2C_1(t_1,t_2)+C_3(t_1,t_2)
\Bigr\},\nonumber\\
&&b_4=
\frac{1}{m}\,\lim_{\tau\to 0}\frac{1}{\tau}
\int_0^\tau \!\!dt_1 \!\!\int_0^\tau \!\!dt_2\, 
C_2(t_1,t_2),\nonumber\\
&&b_5=
\frac{1}{2m}\,\lim_{\tau\to 0}\frac{1}{\tau}
\int_0^\tau \!\!dt_1 \!\!\int_0^\tau \!\!dt_2\!\!
\int_0^\tau \!\!dt_3\, C_4(t_1,t_2,t_3),\nonumber\\
&&b_6=
\frac{1}{4!}\,\lim_{\tau\to 0}\frac{1}{\tau}
\int_0^\tau\!\! dt_1 \!\!\int_0^\tau\!\! dt_2
\!\!\int_0^\tau\!\! dt_3
\!\!\int_0^\tau\!\! dt_4\, C_{5}(t_1,t_2,t_3,t_4).\nonumber
\end{eqnarray}
Here the functions $C_i$ are defined by relations
(\ref{Cs}) and (\ref{Cs2}), and the integrals must be evaluated 
in the coarse-grained limit $\tau\gg \tau_c$.

We did not attempt to give a general prove that $D_4$, with 
coefficients given above,  satisfies the stationary relation $D_4
f_M(p)=0$. Instead, in the next section we evaluate coefficients $b_i$
explicitly for the exactly solvable generalized Rayleigh model.
In this case $D_4$ is found to be the same as for the van Kampen equation
(\ref{VKE}), and therefore the relation $D_4 f_M(p)=0$ is satisfied.

Consider at last the coefficients for the operator $D_2^\star$:
\begin{eqnarray}
&&c_1=\gamma^\star=\left(\frac{\beta}{m}\right)^2\int_0^\infty \!\!dt\,C_0(t)\,
\int_0^\infty \!\!dt\,C_0(t)\,t,
\label{c12}\\
&&c_2=\frac{\beta}{2m}\gamma_{6}=
\frac{\beta}{m}\lim_{\tau\to 0}\frac{1}{\tau}\,
\int_0^\tau \!\!dt_1 \!\int_0^\tau \!\!dt_2\, 
C_6(t_1,t_2).\nonumber
\end{eqnarray}
It is proved in the Appendix that  
\begin{eqnarray}
\frac{c_1}{c_2}=\frac{\beta}{2m}.
\label{relation}
\end{eqnarray}
Because of the factor two in the denominator  
the Maxwellian distribution  $f_M\sim\exp(-\beta
p^2/2m)$ is not stationary for the operator $D_2^\star$,
$D_2^\star f_M(p)\ne 0$. The validity of the relation (\ref{relation})
can be also verified directly, for instance, for $C_0(t)\sim
exp(-t/\tau_c)$.
One might hope that keeping terms of even higher orders  
would restore Maxwellian equilibrium. However, 
the distribution $f_M(p)$ 
does not depend  on $\lambda$, 
which suggests that it must satisfy each term of the $\lambda$-expansion 
separately.

Note that coefficients $c_1$ and $c_2$ are quadratic in cumulants 
$C_0(t)=\langle\!\langle F_0(t_1)F_0(t_2)\rangle\!\rangle$, 
and therefore quadratic in the bath density,
$D_2^\star\sim n^2$. In contrast, the operators
$D_2$ and $D_4$ are
linear in $n$. Therefore, in the low-density limit 
$D_2^\star$ may be neglected,    
and the generalized FPE (\ref{FPE4}) is reduced to the van
Kampen equation (\ref{VKE}). On the other hand, one may expect that
for sufficiently high density of the bath  the operator
$D_4\sim n$ may be 
neglected compared to $D_2^\star\sim n^2$, which
results in a conventional FPE  
\begin{eqnarray}
\frac{\partial f(p,t)}{\partial t}=
\lambda^2 \left\{
A_1\frac{\partial}{\partial p}\,p+
A_2\frac{\partial^2}{\partial p^2}
\right\} f(p,t),
\label{qqq}
\end{eqnarray} 
but with modified coefficients
\begin{eqnarray}
A_1=a_1+\lambda^2c_1, A_2=a_2+\lambda^2 c_2.
\end{eqnarray}
The stationary solution of this  equation is the Maxwellian
distribution  
$\exp\left[-\beta^\star p^2/2m\right]$
with the inverse temperature 
\begin{eqnarray}
\beta^\star=\frac{mA_1}{A_2}=m\,\frac{a_1+\lambda^2c_1}{a_2+\lambda^2c_2},
\end{eqnarray}
which is smaller than that for the bath, $\beta$. Indeed, 
recalling that $a_1/a_2=\beta/m$ and $c_1/c_2=\beta/2m$, one obtains
to order $\lambda^2$
\begin{eqnarray}
\frac{\beta^\star}{\beta}=
1-\lambda^2\,\frac{c_1}{a_1}=
1-\lambda^2\,\frac{\beta}{m}\int_0^\infty dt\,C_0(t)t.
\label{ratio}
\end{eqnarray}
Needless to say, 
the prediction  that the temperature of  a Brownian particle is higher
than the temperature of  the bath  
is in contradiction with basic assumptions of equilibrium statistical
physics and must be subjected to  a thorough scrutiny. 
It might be
instructive to consider a specific model.

\section{Generalized Rayleigh model}
For the generalized Rayleigh model~\cite{piston1} 
it is possible to evaluate the coefficients
in the nonlinear Langevin equation (\ref{MLE}) and  generalized Fokker-Planck 
equation (\ref{FPE4})
analytically. In this  model the bath molecules do not
interact with each other, while 
the Brownian particle interact with molecules not through instantaneous
collisions, as in
the original Rayleigh model~\cite{Kampen_paper,Alkemade,physA},  but 
via a continuous  parabolic repulsive potential. 
Namely,
when the distance between a molecule and the particle $|x_i-X|$
is larger than a given length $R$ the  molecule moves freely. But
when  the molecule enters the ``interaction zone''
$|x_i-X|<R$, it experiences a repulsive parabolic 
potential $\frac{1}{2}k(x_i-\tilde X)^2$, where $\tilde X=X-R$
for a molecule approaching the particle from the left, and
$\tilde X=X+R$ for a molecule from the right. 

As was shown 
in~\cite{piston1}, for this model 
all relevant correlation functions can be evaluated exactly.  
In particular, the correlation functions $C_i(t)$, Eqs. (\ref{Cs}), 
which determine dissipative coefficients in  
the nonlinear Langevin equation (\ref{MLE}), read
\begin{eqnarray}
C_0(t)&=&\omega\,\nu\,p_T^3\,m^{-1}\,\xi_0(\omega t),\label{C123}\\
C_1(t)&=&\omega\,\nu\,p_T\,m\,\,\xi_1(\omega t),\nonumber\\
C_2(t)&=&-\omega\,\nu\,p_T^3\,\,\xi_2(\omega t).\nonumber
\end{eqnarray}
Here  $\omega=\sqrt{k/m}$  is the inverse collision time, 
$\nu$ is the number of molecules per unit length, 
$p_T=\sqrt{m/\beta}$ 
is the thermal momentum of a molecule, and dimensionless functions
$\xi_i(x)$ are
\begin{eqnarray}
\xi_0(x)&=&\frac{2}{\sqrt{2\pi}}\,\theta(\pi-x)
\{\sin x+(\pi-x)\cos x\},\\
\xi_1(x)&=&\frac{1}{\sqrt{2\pi}}\,\theta(\pi-x)
\sin^3 x,\nonumber\\
\xi_2(x)&=&\frac{1}{\sqrt{2\pi}}\,\theta(\pi-x)
\{\sin^3 x+
x(\pi-x)\sin x\}, \nonumber
\end{eqnarray}
where $\theta(x)$ is the Heaviside unit step
function. Substitution of these results into Eqs. (\ref{gammas1})
gives the dissipative constants for 
the nonlinear Langevin equation (\ref{MLE}) 
\begin{eqnarray}
\frac{dp(t)}{dt}=
\lambda^2\gamma_1\,p(t)-{\lambda}^4\gamma_2\,p^3(t)-\lambda\, \xi(t)
\end{eqnarray}  
in the form
\begin{eqnarray}
\gamma_1&=&\gamma_{0}+\lambda^2\delta\gamma+\lambda^2\gamma^\star,\\
\gamma_{0}&=&
\frac{\nu p_T}{m}\int_0^\infty \!\!dx\, \xi_0(x)
=\frac{8}{\sqrt{2\pi}}\,\,\frac{\nu\,p_T}{m},\nonumber\\
\delta\gamma&=&
-\frac{\nu p_T}{m}\int_0^\infty \!\!dx\, 
[2\xi_1(x)+\xi_2(x)]
=-\frac{8}{\sqrt{2\pi}}\,\,\frac{\nu\,p_T}{m},\nonumber\\
\gamma^\star&=&
\frac{\nu^2 p_T^2}{\omega m^2}
\int_0^\infty \!\!dx\,\xi_0(x)
\int_0^\infty \!\!dx\,\xi_0(x)\,x=8\,\frac{\nu^2 p_T^2}{\omega m^2},\nonumber\\
\gamma_2&=&
\frac{\nu}{m\,p_T}\int_0^\infty \!\!dx\,\xi_1(x)=
\frac{4}{3\sqrt{2\pi}}\,\,\frac{\nu}{m\,p_T}.\nonumber
\end{eqnarray}

Consider now the generalized FPE (\ref{FPE4})
\begin{eqnarray}
\frac{\partial f(p,t)}{\partial t}=
\Bigl\{\lambda^2 D_2+\lambda^4 D_4+\lambda^4 D^\star_2
\Bigr\} f(p,t).
\label{qq}
\end{eqnarray}
The coefficients in the operators $D_i$ 
are determined by correlations (\ref{Cs2}). They can be evaluated 
in the same manner as the correlations (\ref{C123}).
Then one can show that the operators $D_2$ and $D_4$ coincide with
those for the original Rayleigh model with instantaneous binary 
collisions and are given  by Eqs. (\ref{D2_binary}) and
(\ref{D4_binary}), respectively. 

The operator $D_2^\star$ originates from non-Markovian
corrections and does not appear in the generalized FPE
for the original Rayleigh model with instantaneous collisions.
It has the form
\begin{eqnarray}
D_2^\star&=&c_1\,\frac{\partial}{\partial p}\,p+
c_2\,\frac{\partial^2}{\partial p^2}\label{c12_rayleigh}
\end{eqnarray}
with coefficients
\begin{eqnarray}
c_1=\gamma^\star=8\,\frac{\nu^2 p_T^2}{\omega m^2},\quad\quad
c_2=16\,\frac{\nu^2 p_T^4}{\omega m^2}.
\label{c12_m}
\end{eqnarray}
In accord with the  general  prediction
(\ref{relation}),  one observes that $c_1/c_2=\beta/2m$, and therefore
the Maxwellian  distribution $f_M(p)=C\exp(-\beta p^2/2m)$ is not
stationary for the operator, $D_2^\star
f_M(p)\ne 0$.

Recall that $D_4$  and $D_2^\star$ scale differently with the molecular
density $\nu$: $D_4\sim\nu$, 
$D_2^\star\sim \nu^2$. This suggests that for
sufficiently high density, the operator $D_4$ can be dropped.  
Inspecting Eqs.(\ref{D4_binary}) and (\ref{c12_m}), one 
finds that the ratio of terms generated by  $D_2^\star$ to those 
produced by $D_4$ is of order 
\begin{eqnarray}
N=\frac{\nu p_T}{\omega m}.
\end{eqnarray}
This is an important parameter of the problem and 
has a meaning of average number of molecules simultaneously
interacting with the particle. For the case 
$N\ll 1$, corresponding to the limit of binary collision,
the operator $D_2^\star$ can be neglected, and the
generalized FPE is reduced to the van Kampen equation (\ref{VKE}). 
The opposite limit $N\gg 1$
corresponds to multiple collisions. In this case the operator
$D_2^\star$ is expected to be more important than $D_4$,
and the generalized FPE to be reduced to the form (\ref{qqq}).
As discussed in the previous section, this equation has a Maxwellian
solution with the inverse temperature $\beta^\star$ smaller than
that for the bath $\beta$. The ratio $\beta^\star/\beta$ is given by
Eq.(\ref{ratio}), which for the given model 
takes the form
\begin{eqnarray}
\frac{\beta^\star}{\beta}=1-\sqrt{2\pi}N\lambda^2. 
\label{pred}
\end{eqnarray}

\begin{figure}
\hfill{}\includegraphics[scale=0.8]{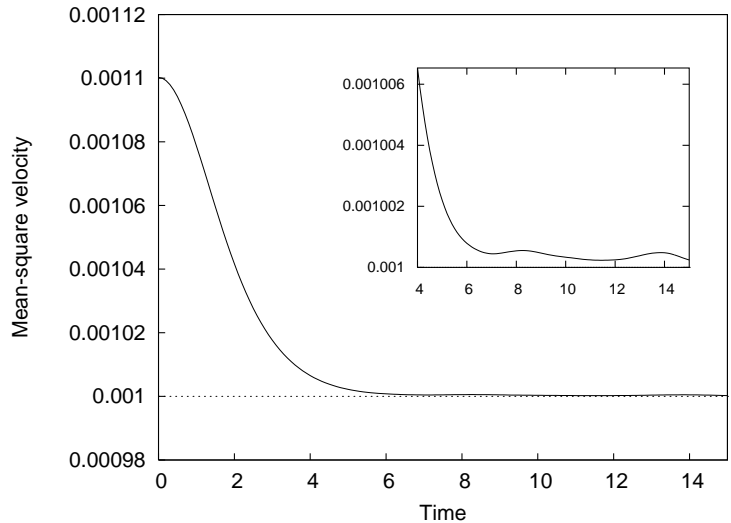}\hfill{}
\caption{ 
The relaxation of the mean-square velocity of a Brownian particle
$\langle V^2(t)\rangle$ for the generalized Rayleigh model. 
Parameters of simulations are 
$\lambda^2=0.001$, $N=100$, $t_{eq}=7.5$. Time is in units of the
collision time $\omega^{-1}$, and velocity is in units of the thermal
velocity  of bath molecules  $v_T$. Zoomed data shown in
the inset suggest that 
the stationary value  exceeds the Maxwellian average 
$\langle V^2\rangle=\lambda^2=0.001$. 
}
\end{figure}
We attempted to verify this prediction in a numerical experiment
using  a simulation  setup 
similar to that described in detail in~\cite{Aaron}. In our
simulation two Poissonian sources of in-going molecules are located
sufficiently far from
the particle to mimic an infinite thermal bath with a Maxwellian velocity
distribution of molecules.  The particle's position  is initially fixed and the
bath is allowed to equilibrate in the field of the fixed particle for
a time $t_{eq}$. 
Then, at the moment $t=0$ the particle 
is released  with an initial velocity $V(0)$, 
and the relaxation of the mean-square velocity $\langle V^2(t)\rangle$ 
towards its stationary value is calculated. 
 Fig. 1 shows the result for the mass ratio
$\lambda^2=10^{-3}$ and the density of the bath corresponding to $N=100$. 
The velocity is in units of the thermal velocity of bath molecules
$v_T=1/\sqrt{m\beta}$, so that according to equilibrium
statistical mechanics $\langle V^2(t)\rangle$ should approach the
equilibrium value $\lambda^2=0.001$. Instead, the stationary mean-square
velocity of the particle appears to be slightly  higher.
However, the exceeding is less than $0.1$ percent which is 
orders of magnitude  lower than predicted by Eq.(\ref{pred}).
The smallness of the alleged deviation from the Maxwellian average  
requires very large number of  
sampling runs, which in our case was about $3\cdot 10^7$.
Also,  the result seems to be sensitive to the value 
of the integration time step $\Delta t$. We used
the velocity Verlet algorithm with $\Delta t\sim 0.001$ in 
units of the collision time $\omega^{-1}$. 
Simulations with larger $\Delta t$ show
no signs of deviations from Maxwellian statistics.

%These features make the simulation very time consuming,  and 
%the result is difficult to interpret unambiguously.
%The deviation can be alternatively attributed to residual fluctuations, or 
%interpreted as a long-lived relaxation tail.
%More accurate numerical results are required to make a distiction.

\section{Concluding remarks} 
In this paper the Fokker-Planck equation
for the Brownian particle is attempted to be derived 
beyond the lowest order in the mass ratio 
$\lambda^2=m/M$. Unlike the approach of van Kampen and Oppenheim~\cite{VKO}, 
the projection operator technique was  
applied to the equation of motion for the particle, rather than to
the Liouville  equation. 
The results seem  both  encouraging and controversial. 

On the one hand, the van Kampen equation, originally designed
for the special case of instantaneous  collisions, is recovered from first
principles for a low-density bath.  
Therefore, in the Markovian limit the method is correct.

On the other hand, it is found that in
general case the equation 
contains the additional operator $\lambda^4 D_2^\star$ 
originating  from non-Markovian corrections which are inevitable
for  any model with finite collision time. 
These corrections are found  
to make the stationary solution non-Maxwellian, 
$D_2^\star f_M(p)\ne 0$,  which is, of course, a very disturbing
result. 
Brownian motion of a massive particle coupled to an infinite
bath in thermal equilibrium, with the Maxwellian velocity 
distribution for bath molecules, 
is often considered as a classical example of 
an ergodic process: all accessible 
microstates are equally probable over a long period of
time, and therefore the equilibrium state is the Gibbs canonical ensemble.
To order $\lambda^2$ this anticipation is supported by solutions of both
Langevin  and Fokker-Planck equations.  
To order $\lambda^4$, the original Rayleigh model  
with instantaneous
collisions leads to the van Kampen equation which also has 
the Maxwellian stationary solution. 
Note that in this last case the ergodic behavior  
can not be deduced from the Central Limit Theorem or 
the assumption of Gaussian noise imposed on the Langevin equation
(which predict the FPE of second order).  
Instead, one can show that
the van Kampen equation is equivalent to    
a  master equation with transition rates 
obeying  the detailed balance condition~\cite{Alkemade}. As well
known, in this case the 
applicability of Boltzmann-Gibbs statistics can be readily   
proved~\cite{Kampen_book}. 
However, whether or not the detailed balance
holds in general case is not known,
and a truly dynamical justification
of Boltzmann-Gibbs statistics 
still seems to be missing~\cite{Cohen}.
A number of 
generalizations of  Boltzmann-Gibbs statistics has been discussed in
recent years, including
those derived directly from underlying dynamics (see for 
instance~\cite{West,Barkai,Hanggi} and references therein).
However, these generalizations usually imply an essential
departure from basic assumptions of Boltzmann-Gibbs  statistical mechanics
such as inelastic collisions, nonextensivity, 
L\'evy statistics for the bath, etc. 
In contrast, the result $D_2^\star f_M(p)\ne 0$
originates merely from  finite duration of 
collisions, which 
suggests deviations from
Boltzmann-Gibbs statistics under a much wider range of 
conditions.

This radical prediction is in contradiction with the results of van Kampen
and Oppenheim~\cite{VKO}. They obtained a generalized FPE in which the
operator $D_2^*$ (in our notations)
has the form of $D_2$ squared, and therefore the stationary solution
is Maxwellian. 
%Our numerical simulation of the generalized Rayleigh model (partly
%reported in~\cite{Aaron})
%also did not give any sign of deviation from the 
%Maxwellian equilibrium. 
This casts some  scepticism about 
the ability of the perturbation approach exploited in this paper
to treat non-Markovian effects. 
The method is based on the
seemingly straightforward
assumption of wide separation of time scales for the particle's
momentum $\tau_p\sim \lambda^{-2}$ 
and for correlation functions of the fluctuating force
$\tau_c\sim \lambda^0$.
However,  
for a finite $\lambda$, the relation $\tau_p\gg \tau_c$ is 
not necessarily true, even if the coupling with slow bath's collective
modes is negligible~\cite{Kapral}.
On the other hand, the van Kampen-Oppenheim approach also relies on
the assumption $\tau_p\gg\tau_c$, and it is not clear why 
the two approaches give different results. Numerical
simulations of the generalized Rayleigh model 
apparently suggest some deviation from the Maxwellian
statistics, but much smaller than the theory 
predicts. It should be stressed, however, that the presented theory is 
asymptotic and implies the weak coupling limit, which is difficult to
approach in a numerical experiment. 
One may hope that further, more accurate  numerical modeling 
would shed some light on these questions. 
%As was discussed in 
%Introduction, there are a number of areas where 
%such progress would be welcome.

\section{Acknowledgment}
%I appreciate valuable comments, often sceptical but never dull, by
%E. Barkai, J.T. Hynes, R. Kapral, and J. Schofield. 
This work was partly supported by NSERC.

\renewcommand{\theequation}{A\arabic{equation}}
\setcounter{equation}{0}
\section*{Appendix}
In this Appendix we prove the relation (\ref{relation}): $c_1/c_2=\beta/2m$.
Using  the definition of $c_1$ and $c_2$, Eq.(\ref{c12}), the relation 
can be written in the form 
\begin{eqnarray}
\int_0^\tau\!\!dt_1\!\int_0^\tau \!\!dt_2\,C_6(t_1,t_2)
\underset{\tau\gg\tau_c}{\longrightarrow}
2\tau\int_0^\infty \!\!dt\, C_0(t)\!\int_0^\infty \!\!dt\, t\, C_0(t).
\label{relation2}
\end{eqnarray}
or taking the derivative
\begin{eqnarray}
\int_0^\tau\!\! dt\, C_6(t,\tau)+
\int_0^\tau \!\!dt\, C_6(\tau,t)
\underset{\tau\gg\tau_c}{\longrightarrow}
2\int_0^\infty \!\!dt\, C_0(t)\!\int_0^\infty \!\!dt\, t\, C_0(t).
\label{relation3}
\end{eqnarray}
Asymptotic relations (\ref{relation2}) and (\ref{relation3}) are
equivalent but the latter is easier to prove. Let us evaluate two
integrals in the left hand side of the relation (\ref{relation3}).

Consider the first integral $I_1=\int_0^\tau dt C_6(t,\tau)$.
Recalling the definition of $C_6$
\begin{eqnarray}
C_6(t_1,t_2)=\int_0^{t_2}dt'\,C_0(t')\int_{t_2-t'}^{t_2} dt''\,
C_0(|t_1-t''|)
\nonumber
\end{eqnarray}
one can write
\begin{eqnarray}
I_1&=&\int_0^\tau \!\!dt\!
\int_0^{\tau}\!\!dt'\,C_0(t')\!
\int_{\tau-t'}^{\tau} \!\!dt''\,C_0(|t-t''|)\nonumber\\
&=&\int_0^{\tau}\!\!dt'\,C_0(t')\!\int_{\tau-t'}^{\tau} \!\!dt''\,
\Bigl\{\int_0^{t''} \!\!dt\, C_0(t''-t)+
\int_{t''}^{\tau} \!\!dt\, C_0(t-t'')\Bigr\}\nonumber\\
&=&\int_0^{\tau}\!\!dt'\,C_0(t')\!\int_{\tau-t'}^{\tau} \!\!dt''\!
\int_0^{t''} \!\!dt''' C_0(t''')+
\int_0^{\tau}\!\!dt'\,C_0(t')\!\int_{\tau-t'}^{\tau} \!\!dt''\!
\int_0^{\tau-t''} \!\!dt''' C_0(t''').
\nonumber
\end{eqnarray}
Introducing the function
\begin{eqnarray}
\Phi(t)=\int_0^t dt' \,C_0(t')
\nonumber
\end{eqnarray}
the above expression can be written as follows
\begin{eqnarray}
I_1=\int_0^{\tau}dt'\,C_0(t')\int_{\tau-t'}^{\tau} dt''\,\Phi(t'')+
\int_0^{\tau}dt'\,C_0(t')\int_{0}^{t'} dt''\,\Phi(t'').
\nonumber
\end{eqnarray}
Noticing that $\Phi'(t)=C_0(t)$, 
it is convenient to integrate the last term by parts:
\begin{eqnarray}
I_1=\int_0^{\tau}dt'\,C_0(t')\int_{\tau-t'}^{\tau} dt''\,\Phi(t'')+
\Phi(\tau)\int_{0}^{\tau} dt\,\Phi(t)-\int_0^\tau dt\,\Phi^2(t).
\label{I1}
\end{eqnarray}

Consider now the second integral in the left hand side of (\ref{relation3})
\begin{eqnarray}
&&I_2=\int_0^\tau dt C_6(\tau,t)=
\int_0^\tau \!\!dt\!
\int_0^{t}dt'\,C_0(t')\int_{t-t'}^{t} \!\!dt''\,
C_0(|\tau-t''|).
\nonumber
\end{eqnarray} 
Since $t''<t<\tau$ the symbol of absolute value may be omitted, and  
 a change of variables  gives
\begin{eqnarray}
I_2\!=\!
\int_0^\tau \!\!dt'\!
\int_{t'}^{\tau}\!\!dt''\,C_0(t''-t')
\int_{t'}^{t''} dt'''C_0(t''')\!=\!
-\int_0^\tau \!\!dt'\, t' \frac{d}{dt'}
\int_{t'}^{\tau}\!\!dt''\,C_0(t''-t')\!\int_{t'}^{t''}\!\!dt'''\,C_0(t''').
\nonumber
\end{eqnarray} 
Evaluation of the derivatives gives
\begin{eqnarray}
I_2\!=\!
\int_0^\tau \!\!dt' t'\,C_0(\tau-t')\int_{t'}^{\tau} \!\!dt''\,C_0(t'')
\!+\!\int_0^\tau \!\!dt'\, t'\,C_0(t')\,\Phi(\tau-t')
\!-\!\int_0^\tau \!\!dt'\, t'\,\int_{t'}^{\tau}\!\!dt''\,C_0(t''-t')C_0(t'').
\nonumber
\end{eqnarray}
Integrating by parts,  the last term in 
this expression can be presented in the form 
\begin{eqnarray}
&&-\int_0^\tau\!\! dt'\, t'\!\int_{t'}^{\tau}\!\!dt''\,C_0(t''-t')C_0(t'')=
-\Phi(\tau)\int_0^\tau \!\!dt'\, \Phi(t')+
\int_{0}^\tau \!\!dt\,\Phi^2(t),\nonumber
\end{eqnarray}
so finally for $I_2$ one obtains
\begin{eqnarray}
I_2\!=\!
\int_0^\tau \!\!dt'\, t'\,C_0(\tau-t')\!\int_{t'}^{\tau} \!\!dt''\,C_0(t'')
+\int_0^\tau \!\!dt\, t\,C_0(t)\,\Phi(\tau-t)
-\Phi(\tau)\int_0^\tau \!\!dt\, \Phi(t)+\int_{0}^\tau \!\!dt \,\Phi^2(t).
\nonumber
\end{eqnarray}

Together with (\ref{I1}) this gives for 
the left hand side of
relation (\ref{relation3})
\begin{eqnarray}
I_1\!+\!I_2\!=\!
\int_0^\tau \!\!dt'\, t'\,C_0(\tau-t')\!\int_{t'}^{\tau} \!\!dt''\,C_0(t'')+
\int_0^{\tau}\!\!dt'\,C_0(t')\!\int_{\tau-t'}^{\tau}\!\! dt''\,\Phi(t'')
+\int_0^\tau \!\!dt\, t\,C_0(t)\,\Phi(\tau-t).
\nonumber
\end{eqnarray}
In the limit $\tau\gg \tau_c$ the first
term in this expression vanishes, while the second and the third terms 
both equal $\Phi(\infty)\int_0^{\infty}dt\,tC_0(t)$:
\begin{eqnarray}
I_1+I_2
\underset{\tau\gg\tau_c}{\longrightarrow}
2\Phi(\infty)\int_0^{\infty}dt\,t\,C_0(t).
\end{eqnarray}
But this is just the right hand side of the relation
(\ref{relation3}), which thus  is proved.

%\end{multicols}

\end{document}